\documentclass[prb,preprint,showpacs,preprintnumbers,amsmath,amssymb]{revtex4}

\usepackage{graphicx}
\usepackage{dcolumn}
\usepackage{bm}

\begin{document}


\title{Effect of c(2x2)-CO overlayer on the phonons of Cu(001): a first principles study}

\author{Marisol Alc{\'a}ntara Ortigoza}
\email{alcantar@physics.ucf.edu}
\affiliation{Department of Physics, University of Central Florida\\
Orlando, Florida 32816, USA\\
and\\
Forschungszentrum Karlsruhe, Institut f\"ur Festk\"orperphysik\\
D-76021 Karlsruhe, Germany
}%

\author{Rolf Heid}
 \email{heid@ifp.fzk.de}
\author{Klaus-Peter Bohnen}%
 \email{bohnen@ifp.fzk.de}
\affiliation{%
Forschungszentrum Karlsruhe, Institut f\"ur Festk\"orperphysik\\
D-76021 Karlsruhe, Germany
}%

\author{Talat S. Rahman}
\email{talat@physics.ucf.edu}
\affiliation{Department of Physics, University of Central Florida\\
Orlando, Florida 32816, USA
}%

\date{\today}

\begin{abstract}
We have examined the effect of a c(2x2) overlayer of CO on the surface phonons of the substrate, Cu(001), by applying the density functional perturbation theory with both the local (LDA) and the generalized-gradient (GGA) density approximations, through the Hedin-Lundqvist and the Perdew-Burke-Ernzerhof functionals, respectively. Our results (GGA) trace the Rayleigh wave softening detected by helium atom scattering (HAS) experiments to changes in the force constants between the substrate surface atoms brought about by CO chemisorption, resolving an ongoing debate on the subject. The calculated surface phonon dispersion curves document the changes in the polarization of some modes and show those of the modes originally along the $\overline{Y}$ direction of the clean surface Brillouin zone (SBZ) which are back-folded along the $\overline{\Delta}$ direction of the chemisorbed SBZ, to be particularly consequential. The vertical and shear horizontal section of $S_1$ in the SBZ of
 the clean surface, for example, is back-folded as a longitudinal-vertical mode, indicating thereby that $S_1$ $-$ predicted a long time back  along $\overline{\Delta}$ for the clean surface $-$ may be indirectly assessed at $\overline{X}$ upon CO adsorption by standard planar scattering techniques. These findings further suggest that some of the energy losses detected by HAS along $\overline{\Delta}$, which were associated to multiphonon excitations of the adlayer frustrated translation mode, may actually correspond to the back-folded substrate surface modes.
\end{abstract}

\pacs{68.35.Ja, 63.20.D-, 68.43.-h, 63.22.-m}
\maketitle

\section{\label{sec:level1}introduction}

Although it is understandable that a large part of theoretical investigations in catalytic surface science is dedicated to developing an understanding of the surface electronic and geometric structure and
energetics of processes such as chemisorption and thermal activation, as a function of catalyst element and surface geometry, two decisive aspects are often overlooked: attestation of the dynamical stability of the model system\cite{bohnen} and understanding of the vibrational dynamics of the reactant-catalyst complex. Clearly, analyses lacking considerations of system dynamics disregard the fact that reaction paths and the so-called \emph{attempt frequencies} or rate preexponential factors pertaining to dynamical processes are themselves determined by the displacement patterns and frequencies of the phonons of the system under consideration, respectively.
Moreover, according to the harmonic transition-state theory, the entire spectrum of phonons
(at the equilibrium and the transition states) of the composite system is required to
determine the attempt frequency of any given process, and not just those vibrational modes in which reactants are primarily involved.
Regardless, it is frequently assumed that the changes in the free energy, which govern the rates of adsorbate processes, involve little
contribution from the substrate because of relatively low chemisorption energies,
as in the case of carbon-monoxide (CO) adsorbed on noble metal surfaces.
The CO binding energy on Cu(001), for example, is 20 times smaller than the carbon-oxygen binding energy and
about three times smaller than the chemisorption energy of CO on Ni(111).\cite{r43, r12} Nevertheless, examination of ultra-violet and X-ray photoelectron spectra signalize a significant interaction between the molecular and the Cu orbitals.\cite{allyn, mariani, umbach, isa, sandell, bjorneholm} Namely, the valence levels of CO molecules adsorbed on Cu are rearranged, broadened, and shifted with respect to those of molecules in the gas phase, in a similar fashion to that observed for Ni(001) $-$ albeit, of course, to a lesser degree.
In each of these cases, molecular adsorption also impacts the substrate phonons. One such example is the case of the hydrogen overlayer on Pt(111), \cite{sampyo,HAS} in which the frequency of the substrate Rayleigh wave at the zone boundaries is modified substantially from its value on clean Pt(111).
Furthermore, even in the case of Cu self-diffusion,\cite{kong} in which the substrate may be expected to play a less important role due to the short-range interaction among coinage metal atoms, it has been shown that the contribution of the substrate to free energy changes in the course of adatom hopping is non negligible. Should the Cu-CO interaction have a range comparable to or longer than the Cu-Cu interaction, for instance, grasping the microscopic details of CO-related reactions facilitated by metallic surfaces will demand taking into consideration all inter-atomic couplings within the system.

We have recently applied the density-functional perturbation theory\cite{r1,r62} (DFPT) to examine the vibrational modes of the c(2x2)-CO adlayer on the Cu(001) surface.  Since all pertinent  interatomic interactions are automatically included in our approach, we were able to calculate the phonon frequencies at
arbitrary propagation directions, i.e., in the entire surface Brillouin zone (SBZ).\cite{ours3} Analysis of the displacement vectors showed that the
CO-modes are influenced by molecule-substrate and molecule-molecule interactions.\cite{ours3} Interactions among neighboring CO molecules, separated by $\sim$~3.6{\AA}, were such as to disperse and/or split the C-O stretch, the frustrated rotation (FR), and the frustrated translation (FT) modes. Interestingly, we also found that omission of the dynamics of the substrate in the calculations lowers the frequency of the Cu-CO stretch mode by $\sim$8 meV with respect
to the value obtained from DFPT calculations. The frequencies of two adlayer modes, the Cu-CO stretch and the FT modes,
were found to depend strongly on whether the local density (LDA) or the generalized gradient (GGA) approximation, as formulated by
Perdew, Burke, and Ernzerhof (PBE), was used.
In fact, the results are a testimony to the unsuitability of LDA to describe the CO adlayer since, contrary to experimental assessments,\cite{allyn, r2}
not only does it render the top adsorption site as a shallow local minimum,\cite{r43} but also implies that the FT mode of the adsorbed
CO molecules is unstable almost everywhere along the $\overline{\Delta}$ and $\overline{\Sigma}$ directions of the c(2x2) SBZ (Fig.~\ref{fig:2a} (b)).
Such dynamical instabilities are indicators of the inability of LDA to describe the Cu-CO interaction which consequently leads to errors in the predicted CO adsorption site on the surface.  The problem, in turn, originates from the expression for the exchange-correlation energy, which is inherently approximated in the
Kohn-Sham formulation and thus leads to a non-exact cancelation of the Coulomb self-interaction.~\cite{kg}
GGA-PBE, though not systematically self-interaction free, introduces an enlargement of the energy gap between the highest occupied orbital (HOMO) $-$ $5\sigma$ $-$ and the lowest unoccupied orbital (LUMO) $-$ $2\pi$* $-$ of CO, thus reducing the hybridization between the $2\pi$* orbital and the metallic $d$-states,~\cite{kre,rap} removing the discrepancy between theory and experiment regarding the preferred adsorption site of CO on Cu(001),~\cite{r43} and accurately reproducing the dispersion of the FT mode as measured by helium atom scattering (HAS) experiments.~\cite{ours3,r2}

In the present work, we turn again to the real-space force constants and the phonon dispersion of c(2x2)-CO/Cu(001) to analyze the effect of CO molecules on the dynamics of the Cu(001) substrate. First of all, the different scenario exhibited by GGA-PBE regarding the C-Cu interaction calls  for revisiting the effect of the CO adlayer on Cu(001). Ellis et al.\cite{r2} observed via HAS measurements that the Rayleigh wave (RW) of Cu(001) softens upon CO adsorption with respect to the clean surface. The later effect was only partially explained by the mass overloading of the CO-\emph{covered} Cu atoms since the softening obtained by simply increasing the mass of such atoms while keeping intact the force constants of the clean surface underestimates the observed softening.
Since no significant changes in the force constants of the Cu substrate was found in the "frozen-phonon" LDA DFT calculations of
Lewis and Rappe\cite{r44} who obtained reasonable agreement with the experiment for the frequency of the \emph{back-folded} RW at $\overline{\Gamma}$, the mass overloading effect was accepted as the main reason for the RW softening. In reality, it is not straightforward to interpret or classify the softening as due to either factor.  On the one hand, while only one of the two Cu surface atoms in the unit cell adsorbs and carries CO, the RW refers (mainly) to the vibration of the first layer which is represented by both atoms in the surface unit cell. On the other hand, \emph{covered} and \emph{bare} atoms relax in opposite directions upon CO chemisorption. In other words, \emph{covered} atoms not only support a CO molecule but also relax outwards, whereas bare atoms relax further inwards with respect to bulk interlayer spacing (GGA-PBE). Developing a rationale for the softening of the RW becomes even more complex given that despite the disagreement between LDA and GGA-PBE in
 the predicted structural features and changes in the force constants of the substrate ~\cite{thesis}, both reproduce reasonably well the HAS measurements of the RW at the zone center.

In this work we also focus on the substrate modes which exist along the $\overline{Y}$ direction of the clean SBZ and are now back-folded along the $\overline{\Delta}$ direction of the chemisorbed SBZ with changed polarization. Our calculations in fact suggest that some of the energy losses detected by HAS in this direction $-$ and ascribed to the multiphonon excitations of the adlayer FT mode {REF} $-$ may actually correspond to surface back-folded modes.  Such backfoldings and polarization changes may be the key for the experimental detection of modes that are otherwise inaccessible to planar scattering techniques. Of particular relevance is the vertical (V - vibration perpendicular to the surface) and shear horizontal (SH - vibration perpendicular to the propagation direction) section of $S_1$ predicted for the clean surface, which may unfold $S_1$ at $\overline{X}$ to planar scattering detection. Specifically, such a mode is back-folded as a longitudinal (L - vibration pa
 rallel to the propagation direction) and vertical mode for the chemisorbed surface and is degenerate with the pure shear-horizontal mode, $S_1$, at the zone boundary.

The rest of this work is organized as follows: Section II contains the computational details. Section III is a summary of results concerning the structure of bulk Cu, the clean Cu(001) surface, and the c(2x2)-CO/Cu(100) chemisorbed surface. In Section IV, we present our results and discussion of the dynamics of all three systems. Finally, Section V contains concluding remarks of this study.

\section{\label{sec:level2}COMPUTATIONAL DETAILS}

Periodic super-cell calculations are performed on the basis of the DFT
formalism and the norm-conserving pseudopotential approach.~\cite{r1} The present results are
derived from the mixed basis (MB) technique.\cite{r52}
 Results using both LDA and GGA  are
 obtained. The former is applied through the Hedin-Lundqvist~\cite{r54}
 parameterization of the exchange-correlation functional, whereas GGA is
introduced via the PBE expression.~\cite{r56}


The clean and the CO-chemisorbed Cu(001) surfaces are simulated with symmetric slabs
inside a tetragonal unit cell containing either 9 (for LDA) or 7 (for GGA-PBE)
layers of Cu. On the chemisorbed surface, CO molecules are
symmetrically located on each side of
the slab. Periodically repeated slabs are separated by a distance equivalent to
11 and 9 vacuum layers, correspondingly.
Integrations inside the Brillouin zone are performed over a discrete mesh of 8x8x8 k-points for bulk Cu and of
8x8x1, and 6x6x1 k-points
for Cu(001) and CO-c(2x2)-Cu(001), respectively.


The calculation of the lattice dynamical matrices at specific q-points of the SBZ is
based on the linear response theory embodied within
DFPT,\cite{r1,r62}
as implemented in the MB scheme in Ref.\onlinecite{r58}.
The dynamical matrices for bulk Cu, Cu(001), and c(2x2)-CO/Cu(001) are
calculated at the q-points of a 4x4x4, a 4x4x1, and a 2x2x1 mesh, respectively.
The real-space force constants in these systems
are obtained by the standard Fourier transform of the
corresponding dynamical matrices.\cite{refagcu}
The force constants of both surfaces are then
matched with those of bulk Cu to model a clean and a
chemisorbed asymmetric slab of 50 layers
and used to obtain the frequencies at
arbitrary q-points.
Surface modes on the clean surface have been identified as those
whose amplitude weight in the two outermost layers is larger than 20$\%$.
On the chemisorbed slab, surface modes and resonances have been identified
as those whose amplitude weight
in the 6 outermost atoms (including C and O) is larger than 20 and 5$\%$, respectively.
For further details of the computational methods, we refer the reader to Ref.\onlinecite{thesis}.


\section{\label{sec:level3} RESULTS AND DISCUSSION OF STRUCTURAL PROPERTIES}

Our results for the bulk Cu lattice parameter (3.57 {\AA} [LDA] and 3.68 {\AA} [GGA-PBE]) are in good
agreement with all-electron\cite{r67} (AE) and pseudopotential\cite{r43,r44} calculations. Nevertheless, our calculated bulk modulus, $B$ (170 GPa [LDA] and 128 GPa [GGA-PBE]), falls
below that provided by AE
calculations.\cite{r67} Discrepancies in this respect are, in fact,
expected since $B$ involves the second derivative of the energy
with respect to the volume, being thus more susceptible
to the differences between AE and PP calculations than the lattice parameter.
As for agreement with
experiment,\cite{r69,r70,r71} LDA underestimates the lattice
constant and yields larger bulk modulus. GGA-PBE overcorrects
LDA, though it reproduces better the experimental bulk modulus
than does LDA. Further details and comparisons can be found in Ref.\onlinecite{thesis}.

A schematic top view of the structure of Cu(001) and the c(2x2)-CO adlayer on Cu(001) is shown
in Fig.~\ref{fig:2a}(a).
The relaxation of the
interlayer distances normal to the surface of the clean Cu(001) has been
extensively studied theoretically.\cite{r1,r7,r43,r44,r48}
Nevertheless, before conducting our
study on the CO-chemisorbed surface, it is essential to test the
applied methodology on the well characterized clean Cu(001) surface; which will also
serve as a reference to appraise
the extent to which CO chemisorption affects it.
Notice in Table~\ref{tab:tab2} that both LDA and GGA-PBE produce
an inwards relaxation of the surface layer of $\sim 3\%$.
These results are in agreement with earlier pseudopotential
calculations\cite{r1,r7,r43,r44,r48} and with
surface structure measurements via the
medium ion energy scattering (MEIS) technique (see Table~\ref{tab:tab2}).\cite{r73}

As illustrated in Fig.~\ref{fig:2a}(a), the primitive super-cell of the CO-covered surface contains two Cu atoms per
layer, which are non-equivalent in odd numbered layers since CO
sits directly above only one of them. Accordingly, atoms in the first layer
are referred either as \emph{covered} or \emph{bare} atoms. To our knowledge, there is no experimental characterization of the substrate
geometry after CO adsorption.
Nevertheless, in agreement with previous calculations,\cite{r43,r44} our results\cite{ours3,thesis}
indicate that the first and third
layers rumple, while the second layer atoms do not.
Quantitative differences, however, arise between LDA and the GGA-PBE results
regarding the interlayer relaxations of the Cu(001) surface upon CO
adsorption.
We refer the reader to Table 1 in Ref.\onlinecite{ours3} for the details of the structure of c(2x2)-CO/Cu(001) and their comparison
with available experimental data and other calculations.\cite{r43,r44}
Here, we only stress that, although both LDA and GGA agree on the fact that CO raises the original inward contraction
of \emph{covered} atoms and even makes them relax slightly outwards with respect to
the bulk situation, such outward
relaxation is two times larger within the GGA-PBE than what it is within the LDA.
In addition, according to LDA, the inward relaxation of bare atoms is slightly decreased by CO, whereas GGA-PBE predicts that bare atoms undergo an
inward relaxation which is even larger ($\sim$ -4.0$\%$) than that of the topmost atoms of the clean surface. Note in passing that for the
Cu-C and C-O bond lengths, for which experimental data is available,\cite{r29}
GGA-PBE gives slightly better agreement to the experimental results than LDA does.\cite{ours3,thesis}
Likewise, our GGA-PBE calculation produces a
chemisorption energy of 0.68 eV at $\theta = 0.5$ ML, to be compared to
the experimental value\cite{r31} of 0.57 eV. LDA, on the other hand, gives a much higher value of 4.00 eV.

\section{RESULTS AND DISCUSSION OF THE LATTICE DYNAMICS}

Just for reference, we mention that our calculated phonon dispersion of
bulk Cu (see Ref.~\onlinecite{thesis}) is in reasonable agreement with neutron inelastic-scattering
measurements (NIS).\cite{r77,r78}
  As for the dispersion of the surface phonons of Cu(001), they have been studied at length recently by DFPT methods
using both LDA and GGA-PBE.\cite{r1,r80} Nevertheless, we have repeated such calculations for consistency in comparison with results for the chemisorbed surface, which is the main subject of this work.
The notation used in this paper for the surface phonons is in accordance with that introduced by Allen et al.\cite{r84}
The (1x1) and c(2x2) SBZ corresponding to the clean and the chemisorbed surfaces, respectively, are shown in Fig.~\ref{fig:2a} (b).
We reiterate that the (1x1) SBZ is defined so that its zone boundaries along the [100] and [110] directions correspond to the $\overline{M}$ and $\overline{X}$ points, respectively. In addition, the $\overline{\Gamma}$-$\overline{M}$, $\overline{\Gamma}$-$\overline{X}$, and $\overline{X}$-$\overline{M}$ segments are denoted as the $\overline{\Sigma}$, $\overline{\Delta}$ and $\overline{Y}$ directions, respectively.
Before proceeding with our analysis, it is important to notice that the q-space of (1x1) SBZ which is outside the c(2x2) SBZ is \emph{back-folded} within the c(2x2) SBZ.

We now turn
briefly to the highlights of the results of our calculations of the phonons of Cu(001).  Full details of our results can be found in Ref.\onlinecite{thesis}.
Note that although a number of surface modes and resonances are found $-$ especially along $\overline{Y}$ $-$,\cite{r84} we describe below only those modes which are
of interest in the discussion of the chemisorbed surface (see Figs.~\ref{fig:5} and ~\ref{fig:7}).
Table~\ref{tab:tab5} summarizes the Cu(001) phonon frequencies at high symmetry q-points ($\overline{X}$ and $\overline{M}$) and compares with those reported in Ref.~\onlinecite{r1} and experiment.\cite{r82,r83}
We should point out that the sharp longitudinal resonance detected in HAS measurements,\cite{r85} is not reproduced in DFPT calculations. An explanation as to why theory and experiment differ on this particular issue remains an open question that deserves further investigation. Nevertheless, for the purpose of this work, we shall not pursue the origin and/or modifications of such longitudinal resonances since they completely vanishe after CO chemisorption.\cite{r2} Notice that the issue of the longitudinal resonance persists for several metal surfaces.\cite{r1} Chis \emph{et al}. have recently addressed the case of Al(001)\cite{chis} and Cu(111).\cite{r87}

With regard  to the c(2x2)-CO/Cu(001) system, as discussed above, LDA indicates that the chemisorption of
CO has little impact on the force constants mediating the
interaction between \emph{bare} atoms and their first NN in the
second layer, whereas those mediating the interaction between
\emph{covered} atoms and their first NN in the second layer, YY,
YZ, ZY, and ZZ, are mildly softened by 12, 12, 24 and 6$\%$,
respectively, as compared to the clean surface.\cite{thesis} GGA-PBE, in contrast,
finds that CO chemisorption modifies the force constants of not
only \emph{covered} atoms but also of \emph{bare} ones.
Naturally, the major effect occurs on \emph{covered} atoms,
whose corresponding force constants are softened by 40, 20, 38,
and 14$\%$, respectively, while those of \emph{bare} atoms are
stiffened by 14, 14, 14, and 9$\%$, respectively. Unlike the
\emph{interlayer} force constants, \emph{intralayer} force
constants between \emph{bare} atoms and their first NN
\emph{covered} atoms are barely altered by CO chemisorption.
Namely, they are $\sim$5$\%$(GGA-PBE) or $\sim$8$\%$(LDA) stiffer
than would be if CO were not present.
Tables~\ref{tab:tab6} and
\ref{tab:tab7}  summarize the frequencies obtained at $\overline{\Gamma}$ and $\overline{X}$,
respectively, and compare them with those found in experiments, when available,
and former theoretical studies. In the sections below, we discuss in detail the
characterization of the
vibrational modes displayed by the Cu(001) substrate and shown in Fig.~\ref{fig:7}. Note, however, that because of the failures of LDA outlined in
Section I and for the sake of clarity, our discussion will be focused on the results obtained via GGA-PBE.  LDA results are mentioned only for cases of substantial discrepancy with GGA-PBE.

\subsubsection{ Substrate modes along $\overline{\Sigma}$}

In the following analysis, one must bear in mind
that modes proper to the clean Cu(001) surface (i.e. within the (1x1) SBZ)
 along the $\overline{\Sigma}$ direction (from the $\overline{\Gamma}$ point
to the $\overline{M}$ point) which appear in the latter half of the same
 are now accessible in the first half, which lies inside the  c(2x2) SBZ (see Fig.~\ref{fig:2a}).

\emph{(a) The $S_1$ mode}. The Rayleigh wave, known as $S_1$,\cite{r84} is the surface mode with the lowest frequency
along $\overline{\Sigma}$ direction. On the clean surface, at  $\overline{M}$,
its polarization is mainly vertical (V) and localized in the first-layer, acquiring an additional longitudinal (L) polarization
as it approaches the $\overline{\Gamma}$-point.
  On the chemisorbed surface, $S_1$ increases its frequency along
$\overline{\Sigma}$, until it reaches the zone boundary and matches, except
for a 1 meV gap, the \emph{back-folded} $S_1$, which has its
maximum at $\overline{\Gamma}$ and decreases its frequency along
$\overline{\Sigma}$.
Regarding the gap between the RW branches at the zone boundary,
we see that the higher (lower) branch
corresponds to a mode whose amplitude weight is primarily
vertical in \emph{bare} atoms (\emph{covered} atoms dragging the
molecule to some extent), involving also a small contribution
from the L-polarized vibration of \emph{covered} (\emph{bare}) atoms.
In fact, the higher branch matches the RW of the clean surface at the zone boundary, suggesting
that \emph{bare} atoms are not affected by the presence of CO.
Nevertheless, we will see that this is not always the case. At $\overline{\Gamma}$, where the softening with respect to the clean
surface is maximal, the \emph{back-folded} $S_1$ mode
corresponds to an \emph{out-of-phase} vibration between
\emph{covered} and \emph{bare} atoms. In this case, the
contribution of \emph{bare} atoms is 50$\%$ larger than that of
\emph{covered} atoms and CO molecules are dragged parallel to
the latter.
Close to $\overline{\Gamma}$, $S_1$ is broadened and appears as a
finite-width resonance whose maximum amplitude weight can be
as low as 6$\%$ in the first two layers. It softens by $\sim$16$\%$ at $\overline{\Gamma}$,
overestimating in fact HAS measurements ($\sim$10$\%$). We observe that although LDA and GGA-PBE predict different effects of
CO chemisorption on the force constants of the first layer, in both
cases $S_1$ softens by the same percentage.

\emph{(b) The $S_2$ mode}. This mode is mostly V-polarized and localized in the second layer at the zone boundary. On the clean surface, this
mode soon forms part of a band of bulk resonances along $\overline{\Sigma}$
whose maximum
amplitude weight comes from either the second and first layers (with V- and L- polarization,
respectively) when close to the SBZ boundary, or first
layer (with mixed V-L polarization) when close to $\overline{\Gamma}$. GGA-PBE and LDA yield slightly different results as to its prevalence along $\overline{\Sigma}$ and its degree of localization (see Ref.\onlinecite{thesis}).

On the chemisorbed surface, however, $S_2$ only appears \emph{back-folded}
from $\overline{M}$ to $\overline{\Gamma}$. It steeply disappears as
soon as it immerses into the bulk band. At $\overline{\Gamma}$, $S_2$ is
well inside the bulk band, even though, no coupling to bulk modes
is observed. On the contrary, it is more localized on the
chemisorbed surface than on the clean surface (at $\overline{M}$). $S_2$, in addition, stiffens by $\sim$~3 meV on the chemisorbed surface.

\emph{(c) The $L_1$ mode and the corresponding shear-horizontal (SH) branch}.
$L_1$ refers to a L-polarized mode mostly localized in the first layer of the clean surface at the zone boundary of the SBZ. It rapidly becomes a resonance as it approaches $\overline{\Gamma}$. As a consequence, on the
CO-covered surface, it appears mainly as a \emph{back-folded} mode.
At $\overline{\Gamma}$, the \emph{back-folded} $L_1$ and its degenerate SH
pair stiffen very slightly ($\sim$~1$\%$) due to the chemisorption. Neither $L_1$ nor its degenerate SH
pair are totally localized in the first or second
 layers (as in the clean surface) but exhibit a significant contribution of $\sim$60$\%$ to the amplitude
weight from deeper layers. Outside $\overline{\Gamma}$, the \emph{back-folded}
$L_1$ rapidly disappears, while the \emph{back-folded} SH mode
disperses and becomes more localized towards the zone boundary, where it matches the branch that originally
appears on the clean surface but vanishes rapidly back to
$\overline{\Gamma}$.

\subsubsection{Substrate modes along $\overline{\Delta}$}

Here, one should notice that, like the \emph{back-folding} along $\Sigma$, modes proper to the clean surface along $\overline{Y}$ are
\emph{\emph{back-folded}} to $\overline{\Delta}$ in the c(2x2) SBZ (see Fig.~\ref{fig:2a} (b)).

\emph{(a) $S_1$ and the corresponding L-branch}.
$S_1$ is the mode of lowest energy of Cu(001) and is totally localized in the first layer at the $\overline{X}$ point.
It cannot be detected by standard planar scattering techniques on the clean surface
since its polarization is SH all along $\overline{\Delta}$. $S_1$
changes rapidly to V- polarization along $Y$ but, to our knowledge, no experimental
data along $\overline{Y}$ is yet available.

On the chemisorbed surface, $S_1$ is totally localized in the first layer and the CO overlayer.
The latter vibrates \emph{in-phase} with the
\emph{covered} Cu atoms but with much smaller amplitude.
$S_1$ softens at $\overline{X}$ by $\sim$8$\%$ (see Tables~\ref{tab:tab5} and ~\ref{tab:tab7}). Interestingly, the calculations reveal  a L-polarized mode that is  degenerate with $S_1$ at $\overline{X}$, albeit the degeneracy is broken outside $\overline{X}$. Such mode
corresponds to the section of $S_1$ along $\overline{Y}$ in the SBZ of the clean surface, which is
\emph{back-folded} along $\overline{\Delta}$ on the
chemisorbed SBZ and changes polarization from SH to L.
The \emph{back-folded} $S_1$ remains L-polarized as it goes across
$\overline{\Delta}$ towards $\overline{\Gamma}$ up to the crossing point with the RW,
where it becomes V-polarized. It is slightly softened around the zone
boundary, just as much as $S_1$ at $\overline{X}$; nonetheless, the softening
becomes stronger $-$ similar to that of the RW $-$ right after crossing
the RW and the transitioning to V-polarization.
Note that Ellis et al. observed some
 HAS peaks precisely at the region where \emph{back-folded}
 $S_1$ is V-polarized.\cite{r2} Those peaks were at the time said to be associated with
 multi-phonon excitation bands. The
 excellent fit of their measured dispersion to our \emph{back-folded}
 $S_1$ ($\overline{Y}$ is
 \emph{back-folded} onto $\overline{\Delta}$), however, suggests that this latter mode is observed rather than multi-phonon excitation bands.

\emph{(b) The $S_4$ mode}. This is the RW along $\overline{\Delta}$.
On the clean surface, it is essentially V-polarized, although a
L-contribution is also present.
Although its amplitude weight is greatest at the first layer,
it decays slowly as a function of the layer depth.
On the chemisorbed surface, it is also
mostly localized ($\sim$60-70$\%$) in the first layer and, to
lesser degree, in the molecule. $S_4$ also displays a small splitting at $\overline{X}$ (see Table~\ref{tab:tab7}), yet in this case both branches
are softened with respect to that of the clean surface by 9.6 and
15.0$\%$ at $\overline{X}$, while HAS measurements\cite{r2} find
 a softening of $\sim$8.2$\%$. From the first
layer, only the \emph{covered} (\emph{bare}) atoms contribute to the
mode with lower (higher) energy.
The \emph{back-folded} $S_4$, on the other hand,
originally arises along $\overline{Y}$ on the clean surface and has V-polarization
at the zone boundary and a predominantly SH-character as it crosses $\overline{Y}$. On the chemisorbed surface, this mode is also
V-polarized close to the zone boundary. In GGA-PBE, it broadens and
becomes a resonance as soon as it immerses into the bulk band,
reappearing as a surface mode close to $\overline{\Gamma}$
with mixed L- and SH-polarization. In LDA, however, $S_4$ remains
highly localized on the surface and
changes smoothly to L-polarization.~\cite{thesis}

\emph{(c) The $S_5$ mode}. The polarization of this mode is SH and it is predominantly localized in the second layer on the clean surface. On the chemisorbed surface, it is also strongly localized in the second layer
but with mixed SH- and L-polarization. It
rapidly becomes a resonance along $\overline{\Delta}$ towards $\overline{\Gamma}$.
At $\overline{X}$, $S_5$ splits (see Table~\ref{tab:tab7}). In GGA-PBE
both branches are totally localized in the second layer. One of these
softens by 8.4$\%$ and the other stiffens by 3.5$\%$.
LDA shows that the lower branch
softens while the other - bearing $\sim$25$\%$
contribution from deeper layers - does not change at all.

\emph{(d) The $S_6$ mode}. This mode developes inside the spectrum gap, close to the zone boundary with a frequency of 23.6 meV (see Table~\ref{tab:tab7}). On the clean surface, it has a predominant L-polarization in the first layer.
On the chemisorbed surface, however, it is found to be degenerate with a
SH pair at $\overline{X}$.
Note in Fig.~\ref{fig:7} that
the degeneracy is broken inside the SBZ.
Our calculations find that $S_6$ - and the SH-branch -  remain nearly
dispersionless and involve
an additional V polarized second layer vibration, as found on the clean
surface.
These modes are, incidentally, more localized
in the first layer of the chemisorbed surface than in that of clean Cu(001).
At the right end of the bulk band gap (see Fig.~\ref{fig:7}), $S_6$ slightly softens and becomes a
resonance while it penetrates the bulk band. The SH branch, in contrast,
extends well inside the bulk band and becomes
more localized at the top two layers. According to GGA-PBE, $S_6$ softens at $\overline{X}$ by 2.0$\%$,
while LDA predicts no softening (see Table~\ref{tab:tab7}).

\section{SUMMARY}

A first-principles study of the dynamics of a c(2x2)CO overlayer chemisorbed  on Cu(100) has been presented. Our calculations show that LDA displays a mild effect of CO chemisorption on the relaxations
and force constants of the surface. The LDA predicted adsorption site is also not in agreement with experimental results.
Nevertherless, LDA is able to give good agreement with the HAS data for the frequency at the zone center of
the \emph{back-folded}
RW. This feature is rather unexpected in
consideration of the poor ability of LDA to describe the Cu-CO
interaction and its less successful description of the acoustic
modes of bulk Cu and the RW in Cu(001).\cite{thesis} GGA-PBE, on the other
hand, finds that CO chemisorption significantly perturbs the
structure and the first NN force constants of the surface layer
atoms. Softening of the RW is well reproduced; only slightly
overestimated by $\sim$1 meV at $\overline{\Gamma}$. It is surprising that,
while LDA and GGA-PBE differ considerably in the response of the substrate
to CO chemisorption, the actual percentage softening of
the \emph{back-folded} RW with respect to the clean surface is
comparable.
We find that softening of the RW along $\overline{\Sigma}$ and $\overline{\Delta}$ is
not necessarily connected with the vibration of the \emph{covered}
atoms, which indicates that mass overloading alone cannot account for it.
If that were to be the case, all surface modes involving the first layer atoms
would be softer. On the contrary, $L_1$, for example, does not
soften, notwithstanding the leading involvement of first layer
atoms. In fact, it slightly stiffens, consistent with the
hardening of intralayer force constants of the first layer. Moreover,
softening/stiffening of the various modes cannot be simply dictated by the
atomic layer, propagation direction, and polarization characteristics,
indicative of complexity in the modifications in Cu force constants.
For example, $S_2$, the V-mode in the second layer, stiffens.
In turn, while the $S_1$ mode (along $\overline{\Delta}$)
slightly softens in the region where it is L-polarized and totally
localized in the first layer, it
undergoes a stronger softening in the region where it is
V-polarized and involves contributions from deeper layers.
Softening of the RW seems thus due
not only to the mass of CO but also to
 interactions of longer range, i.e. beyond first NN and involving
 deeper layers, that subdue differences in the local bonding of
 surface atoms and result in an overall softening of the RW, which is
 sometimes independent on whether \emph{covered} or
 \emph{bare} atoms are primarily involved.

Our results call attention to the importance of the folding of  $\overline{Y}$ (of the 1x1
 SBZ) onto $\overline{\Delta}$ (of the c(2x2) SBZ), which displays
 \emph{back-folded} modes along the latter. For example,
 \emph{back-folded} $S_1$ and \emph{back-folded} $S_4$ are
 found along $\overline{\Delta}$ with changed polarization that may make
them observable by standard scattering techniques. In particular, $S_1$ acquires V-polarization
 close to $\overline{\Gamma}$ and L-polarization close to $\overline{X}$.
The good agreement for the dispersion of this latter mode
HAS measurements \cite{r2} is convincing evidence that these peaks arise from the \emph{back-folded} $S_1$  mode, rather than from multi-phonon excitation bands.
Perhaps even more importantly, \emph{back-folded} $S_1$ becomes as well
 discernible to planar scattering spectroscopy techniques close to the zone
 boundary since its polarization changes from SH- (along
 $\overline{Y}$ of the (1x1) SBZ) to L- polarization (along
 $\overline{\Delta}$ of the c(2x2) SBZ) in that region. The above thus implies
 that the frequency of $S_1$ (SH-polarized in Cu(001)) can indirectly be
 measured at $\overline{X}$ via \emph{back-folded} $S_1$. We believe, in fact, that
 some of the peaks observed and assigned to \emph{2T overtones}\cite{r2} peaks by Ellis \emph{et al}.
 along $\overline{\Delta}$ may instead correspond to \emph{back-folded}
$S_1$, in the region where is L-polarized. Our results here thus call for a new interpretation of the HAS data.

\begin{acknowledgments}
This work was supported in part by grant CHE-0741423 from NSF-USA. Computations
were performed at the Institut f\"ur Festk\"orperphysik, Forschungszentrum Karlsruhe, Germany.
Marisol Alc{\'a}ntara Ortigoza is thankful to the Forschungszentrum Karlsruhe for
financial support during her stays in Karlsruhe.

\end{acknowledgments}

\bibliography{prb}

\newpage

\begin{table}
\caption{\label{tab:tab2}
Percentage change of the interlayer spacing of the outermost layers of Cu(100) compared to the bulk situation.
}
\begin{ruledtabular}
\begin{tabular}{|c|c|c|c|c|}
\hline
&\multicolumn{2}{c|}{Theory\footnotemark[1]} & \multicolumn{2}{c|}{Experiment}\\
\hline
 &LDA & GGA-PBE&SPLEED\footnotemark[2] &MEIS\footnotemark[3] \\
\hline
$\Delta d_{12}$&-2.57&-2.82&-1.2&-2.4 \\
\hline
$\Delta d_{23}$&+0.55&+0.58&+0.9&+1.0\\
\hline
$\Delta d_{34}$&+0.30&fixed&-&-\\
\hline
\end{tabular}
\end{ruledtabular}
\footnotemark[1]{ This work }
\footnotemark[2]{ Ref.~\onlinecite{r72} }
\footnotemark[3]{ Ref.~\onlinecite{r73} }
\end{table}

\begin{table*}
\caption{\label{tab:tab5}
Frequencies (in meV) of the surface modes of Cu(100)
at the high symmetry points $\overline{X}$ and $\overline{M}$ (see Fig.~\ref{fig:2a} (b)).
The main polarization (SH, L, or V) is denoted in parenthesis and
the superscript indicates the layer showing largest amplitude weight.
}
\begin{ruledtabular}
\begin{tabular}{|c|c|c|c|c|c|c|}
\hline
 & &\multicolumn{3}{c|}{Theory}& \multicolumn{2}{c|}{Experiment} \\
\hline
 & &\multicolumn{2}{c|}{LDA} &GGA-PBE& EELS & EELS\\
\hline
 & &This work&Ref.~\onlinecite{r1}&This work&Ref.~\onlinecite{r82}&Ref.~\onlinecite{r83} \\
\hline
$\overline{X}$&$S_1 (SH^1)$&9.7&9.9&9.1&-&-\\
\hline
 &$S_4 (V^1)$&14.6&14.0&13.5&13.2&-\\
\hline
 &$S_5 (SH^2)$&15.6 &15.0 &14.3 &- &- \\
\hline
 &$S_6 (L^1)$&27.0 &26.1 &24.1 &- &25.2 \\
\hline
$\overline{M}$&$S_1 (V^1)$&18.7 &17.9 &16.9 &17.0 &- \\
\hline
 &$S_2 (V^2)$&22.3&- &20.0 &- &20.3 \\
\hline
 &$(SH^{1,2})$&22.3 &21.1-21.7 &20.1 &- &- \\
\hline
 &$L_1(L^{1,2})$&22.3 &- &20.1 &- &~20.4 \\
\hline
\end{tabular}
\end{ruledtabular}
\end{table*}

\begin{table}
\caption{\label{tab:tab6}
Frequencies (in meV) at $\overline{\Gamma}$ of the surface vibrational modes of c(2x2)-CO/Cu(001).
}
\begin{ruledtabular}
\begin{tabular}{|c|c|c|c|c|}
\hline
&\multicolumn{3}{c|}{Theory} & Exp.\\
\hline
&\multicolumn{2}{c|}{LDA} &\multicolumn{1}{c|}{GGA} &HAS \\
\hline
 &This work\footnotemark[1]&Ref.~\onlinecite{r44}\footnotemark[2]&This work\footnotemark[1]&Ref.~\onlinecite{r2}\\
\hline
$S_1$& 15.8 & 16.0 &14.2& 15.2\\
\hline
$L_1$& 22.6 & -   & 20.2 & -\\
\hline
$S_2$& 23.1 & 23.2& 20.4& -\\
\hline
\end{tabular}
\end{ruledtabular}
\footnotemark[1]{DFPT.}
\footnotemark[2]{DFT-FD.}
\end{table}

\newpage

\begin{table}
\caption{\label{tab:tab7}
Frequencies (in meV) at $\overline{X}$ of the surface vibrational modes of c(2x2)-CO/Cu(001).
}
\begin{ruledtabular}
\begin{tabular}{|c|c|c|c|}
\hline
&\multicolumn{2}{c|}{Theory} &Experiment\\
\hline
 &LDA & GGA & HAS \\
\hline
 &This work&This work&Ref.~\onlinecite{r2}\\
\hline
$S_1$&9.3&8.3&-\\
\hline
$S_4$&13.1&11.6&12.3\\
     &13.3&12.2&    \\
\hline
$S_5$&15.2&13.1&-\\
     &15.6&14.8& \\
\hline
$S_6$&27.1&23.6&-\\
\hline
\end{tabular}
\end{ruledtabular}
\end{table}

\clearpage

\begin{figure}
{\huge (a)}
\vskip4pt\includegraphics[width=0.4\textwidth]{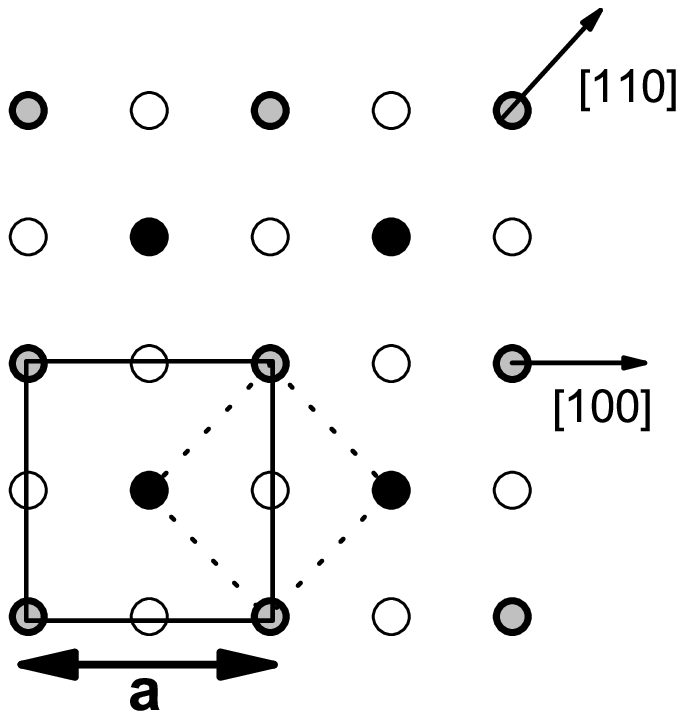}{\tiny . \\ \huge (b) \\}
\vskip 0.29in minus -0.5in
\includegraphics[angle=0, width=0.4\textwidth]{ARHBfig_2}
\caption{\label{fig:2a}
}
\end{figure}

FIG.~\ref{fig:2a}: (a) The top view of the surface shows CO (grey circles),
and first (filled circles) and second (open circles) layer atoms of Cu(100).
The 1x1 (dashed line) and the c(2x2) (solid line) surface unit cells are underlined.
(b) The corresponding (1x1) (dotted line)  and c(2x2) surface Brillouin zones (solid line)
showing the $\overline{\Gamma}$, $\overline{X}$, and $\overline{M}$ points; and the $\overline{\Delta}$, $\overline{\Sigma}$, and $\overline{Y}$ directions.

\begin{figure}
\includegraphics[angle=0, width=0.467\textwidth]{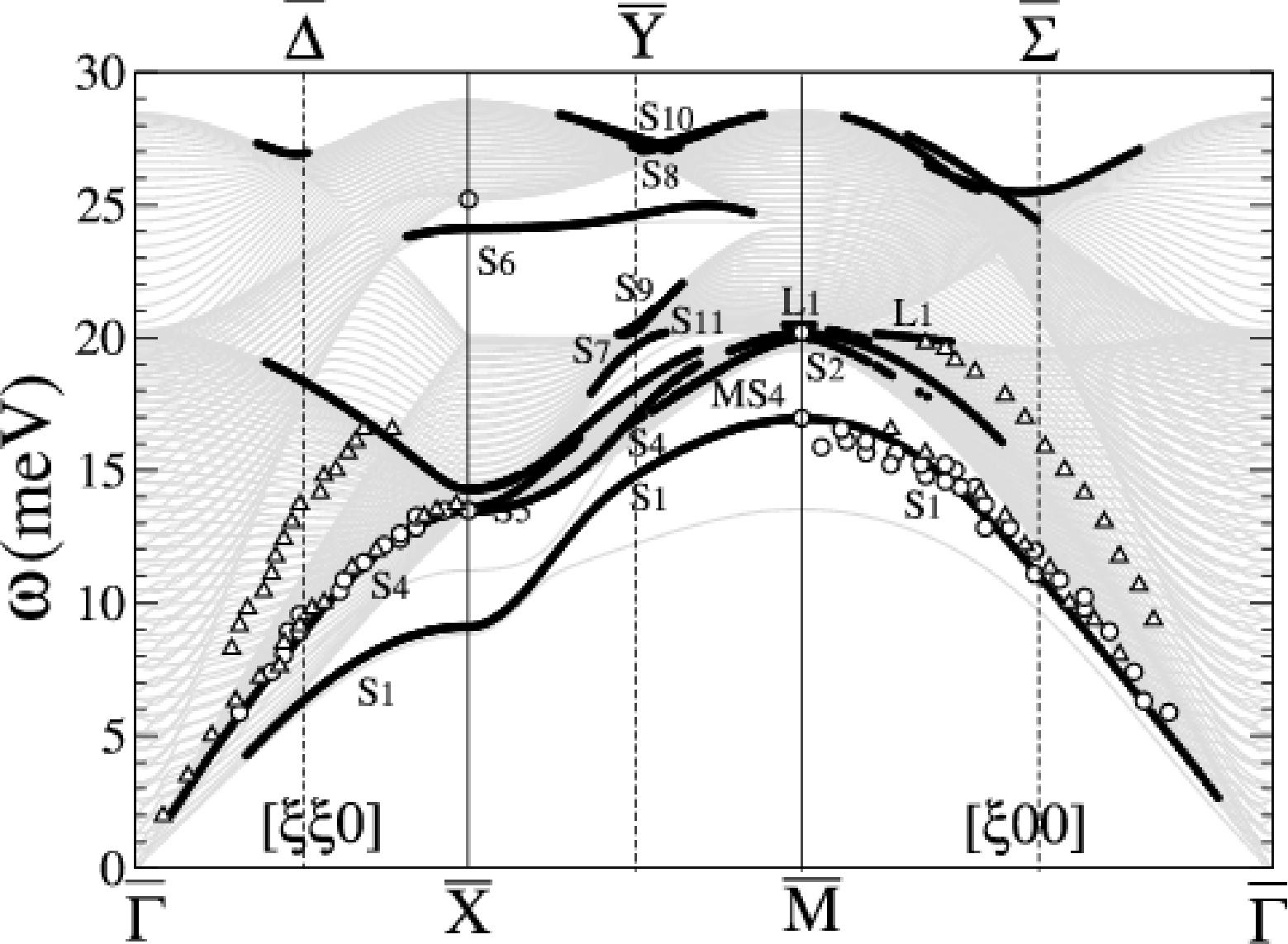}
\caption{\label{fig:5}
}
\end{figure}

FIG.~\ref{fig:5}: GGA-PBE phonon dispersion of Cu(100), modelled by a 50-layer slab.
Theoretical surface modes (filled circles) are compared with HAS (open
triangles) and EELS (open circles) measurements taken from
Refs.~\onlinecite{r81,r82,r83}.

\begin{figure}
\includegraphics[angle=0, width=0.448\textwidth]{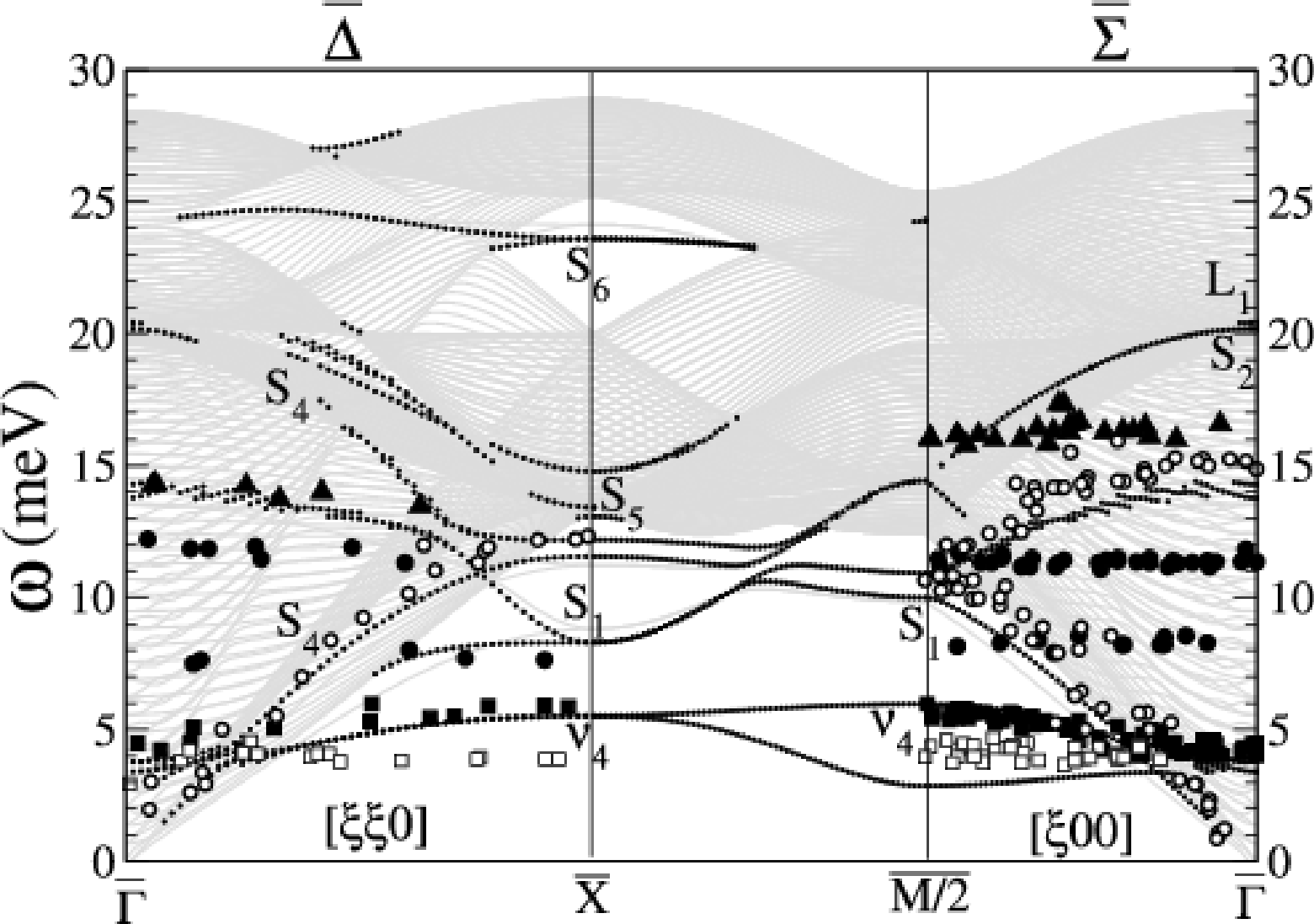}
\caption{\label{fig:7}
}
\end{figure}

FIG.~\ref{fig:7} : GGA-PBE phonon dispersion of c(2x2)-CO/Cu(100), modelled by
a Cu 50-layer slab. Note that the dispersion of the high-lying modes, which correspond to the C-O stretch and the Cu-CO stretch modes, are omitted in this figure since the emphasis here is on the modes of the substrate. Filled circles denote theoretical surface modes.
Experimental data are taken from Ref.~\onlinecite{r2}: Filled circles and triangles
were associated with multiphonon processes. Open circles correspond to the substrate Rayleigh wave.
Squares were associated with the FT mode of CO on the perfect c(2x2)
structure (filled) and on defects in the adlayer at lower coverage (open).

\clearpage

\end{document}